\documentstyle[12pt]{article}
\begin{document}
\tolerance 6000
\hbadness 60007

\title{A FREE RELATIVISTIC ANYON WITH CANONICAL SPIN ALGEBRA}
\author{D. Dalmazi\thanks{e-mail:
dalmazi@grt$\emptyset\emptyset\emptyset$.uesp.ansp.br}~~
and A. de Souza Dutra\thanks{e-mail
:dutra@grt$\emptyset\emptyset\emptyset$.uesp.ansp.br} \
\\UNESP - Campus de Guaratinguet\'a -
DFQ \\Av. Dr. Ariberto Pereira da Cunha, 333 - CEP 12500-000 \\
Guaratinguet\'a - SP - Brasil}
\date{}
\maketitle
\thispagestyle{empty}
\begin{abstract}

We discuss a relativistic free particle with fractional spin in 2+1
dimensions, where the dual spin components satisfy the canonical
angular
momentum algebra $\left\{ S_\mu , S_\nu \right\}\,=\,\epsilon_{\mu
\nu
\gamma}S^\gamma $. It is shown that it is a general consequence of
these features that the Poincar\`e invariance is broken down to the
Lorentz one, so indicating that it is not possible to keep
simultaneously
the free nature of the anyon and the translational invariance.
\end{abstract}

\newpage

\section{Introduction}

In the last few years the study of particles with generalized
statistics,
the so called anyons \cite{Wilczek}, has been a growing field of
interest. This is mainly due to the possibility of explaining
very important physical phenomena, the fractional quantum Hall effect
and
the high temperature superconductors \cite{hall,super}.

However, most of the work on this matter was done through
nonrelativistic
models, or by using the statistical field
$A_\mu$ of the topological Chern-Simons term \cite{Gamboa?}.
Notwithstanding, some authors have been looking for the answer to the
question whether this statistical field not only changes the
statistics,
but also endows the anyons with interaction.  One has been
looking in particular, into the possibility of constructing a
point-particle model for
anyons, without the need of using any statistical field
\cite{Japa,Jackiw,Forte,Plyuschay}. This quest should necessarily
begin
by studying the representations of the Lorentz group in 2+1
dimensions
\cite{Japa,Jackiw,Forte}.

The Lorentz group in 2+1 dimensions corresponds to the SO(2,1) group
whose irreducible unitary representations, are all labeled by a set
of
half-integer and integer numbers
\cite{Japa}. Thus if we are interested in a Lorentz covariant
description of particles with fractional spin, we have to deal in
principle with an infinite dimensional representation of the Lorentz
generators. On the other hand, Jackiw and Nair have shown in
\cite{Jackiw}
that the anyon posses just one physical polarization state  for a
given
sign of the energy, and they have obtained an induced representation
for
the Poincar\`e group on those physical polarization states. What is
actually important for us is that, in the induced representation, the
spin components commute ($\left\lbrack S_\mu ,
S_\nu\right\rbrack\,=\,0$)
and therefore do not satisfy the usual angular momentum algebra.
Nevertheless, some point-particle models corresponding to the induced
representation have appeared in the literature
\cite{Forte,Plyuschay,Chaichian}. However, as far as we know, none of
the models discussed up to now has been able to get two ingredients
holding simultaneously, namely the free nature of the relativistic
anyon
and the canonical spin algebra $\left\{ S_\mu ,
S_\nu\right\}\,=\,\epsilon_{\mu \nu \gamma}S^\gamma$ for the spin
components \cite{Plyuschay}. In this work we
intend to explore the consequences of
imposing these features simultaneously.
In the next section we show explicitly how this can be carried
through in
a particular point-particle model. In section three, we perform a
model
independent analysis, then in section four we draw some conclusions
and summarize the work.

\begin{sloppypar}
\section{A model for a free particle with fractional spin}
\end{sloppypar}
The Poincar\`e algebra in 2+1 dimensions is given by:

\begin{eqnarray}
\bigl\lbrack J_\mu , J_\nu \bigr\rbrack \,&=&\,\epsilon_{\mu \nu
\gamma}
J^\gamma \nonumber\\
\bigl\lbrack J_{\mu} , P_\nu \bigr\rbrack
\,&=&\,\epsilon_{\mu
\nu \gamma} P^\gamma \\
\bigl\lbrack P_\mu , P_\nu \bigr\rbrack
\,&=&\,0\nonumber,
\end{eqnarray}
where $J_\mu$ are the dual components of the total angular momentum
tensor, {\it i.e.} $J_\mu \,=\,\epsilon_{\mu \nu \gamma}J^{\nu
\gamma}$
and we are using $\eta_{\mu \nu}\,=\,diag\,(-,+,+)$,
$\epsilon_{012}\,=\,+1$.

Let us suppose now that we have a relativistic
model for a point-particle whose total angular momentum tensor and
momenta components are given in the phase space by

\begin{equation}
J_{\mu \nu}\,=\,x_\mu p_\nu \,-\, p_\mu x_\nu \,+\, n_\mu \pi_\nu \,-
\,
\pi_\mu n_\nu\,, \; P_\mu \,=\, p_\mu.
\end{equation}

\noindent Assuming that the vectors $x_\mu ,p_\mu , n_\mu ,\pi_\mu$
satisfy the canonical Poisson brackets, $\left\{ x_{\mu} ,
p_{\nu} \right\} \,= \, \eta_{\mu \nu} \, = \, \left\{ n_\mu ,
\pi_{\nu}
\right\}$, with the
remaining brackets vanishing, it is easy to derive the Poincar\`e
algebra
in terms Poisson brackets. Note that the variables $\left( n_\mu ,
\pi_\mu\right)$ are just introduced to describe the spin of the
particle
$S_\mu \,=\,\epsilon_{\mu \nu \beta}n^\nu \pi^\beta$, which
satisfies $\left\{ S_\mu , S_\nu \right\}\,=\,\epsilon_{\mu \nu
\gamma}S^\gamma$. The Poincar\`e algebra (1) possess two quadratic
Casimir
invariants which can be represented in our model by $p^2$ and $S\cdot
p$,
they will be used in the form of constraints to specify the mass and
helicity of the particle respectively:

\begin{eqnarray}
\phi_1 \,&=&\, p^2\,+\,m^2 \,\approx \, 0 \\
\phi_2 \,&=&\, S\cdot p \,+\, \alpha m \, \approx \, 0 \quad ,
\end{eqnarray}

\noindent where $\alpha$ is an arbitrary real constant. In analogous
fashion to the $\alpha \,=\, 1/2$ case \cite{Jackiw}, Eq.(4) will
play
the role (for arbitrary $\alpha$) of the Dirac equation
\cite{Russos,Jackiw}. We expect that the constraints $\phi_1 ,
\phi_2$ be first class, in order to specify the physical states which
belong to the Poincar\`e representation specified by $\alpha$ and
$m$,
such that the hamiltonian of the model can be assumed to be of the
form

\begin{equation}
H\,=\,e\,\phi_1 \,+\, \sigma \,\phi_2 \,+\,\lambda^a \chi_a,
\end{equation}

\noindent where $e,\,\sigma$ and $\lambda^a$ ($a$\,=\,1,...,4) are
the
lagrange multipliers, whereas $\chi_a$ represent four second class
constraints. The reason why we need exactly four second class
constraints
comes from a simple counting of degrees of freedom as follows; by
assuming that the spin components $S_\mu$ and the momenta $p_\mu$ are
parallel, one can show for arbitrary $\alpha$ that $g\,=\,2$, where
$g$
is the gyromagnetic constant \cite{Poli}. This result agrees with
previous calculations \cite{Gamboa} in field theory.
If this is indeed the case,
the spin of a particle in 2+1 dimensions is
entirely
specified by the helicity $S \cdot p$ which is fixed in our model
by the constraint
$\phi_2$ and since $\phi_2$ is assumed to be
a first class constraint, it will eliminate two degrees of freedom of
the extra six variables
$\left( n_\mu , \pi_\mu \right)$ introduced to describe the spin
degrees
of freedom. So we have to further impose four second class
constraints in
order to eliminate the remaining variables. It can be checked that
the
easiest way to assure that $S_\mu$ and $p_\mu$ be parallel is to
impose
the constraints \cite{Plyuschay,Chaichian}

\begin{equation}
\chi_1 \,=\, p \cdot n \,= \, 0 \,, \;\; \chi_2 \,=\, p \cdot \pi
\, =\, 0 \quad ,
\end{equation}

\noindent these constraints imply

\begin{equation}
S_\mu \,= \, \frac{\left(S \cdot
p\right)}{p^2}\, p_\mu \quad .
\end{equation}

\noindent Due to the fact that (6) does not depend on
$x_\mu$, after quantization the commutation relations of $p_\mu$
would
not be changed by such constraints and we would have $\left\lbrack
S_\mu
, S_\nu \right\rbrack \,=\, 0$ as a consequence of (7). In order to
recover the usual spin algebra one must further impose $x_\mu$
dependent constraints, for instance,

\begin{equation}
\chi_3 \,=\, x\cdot n\,=\,0,\;\,\chi_4\,=\,x\cdot \pi\,=\,0\quad .
\end{equation}

\noindent The constraints (6) and
(8) form a set of four independent second class
constraints which can be eliminated by a Dirac bracket
$\left\{ \, ,\, \right\}_D$ leading to

\begin{eqnarray}
\left\{ \pi_\mu , \pi_\nu \right\}_D \,&=&\, 0 \,=\,
\left\{ n_\mu , n_\nu \right\}_D \, , \nonumber \\
\left\{n_\mu , \pi_\nu \right\}_D \,&=&\, \eta_{\mu \nu} \nonumber \\
\left\{ p_\mu , n_\nu \right\}_D \,&=&\,
\frac {\left[
(\pi \cdot n)n_\mu \,-\, n^2\pi_\mu\right]p_\nu }{S^2} \, ,\nonumber
\\
\left\{p_\mu , \pi_\nu \right\}_D \,&=&\, \frac {\left[\pi^2n_\mu
\,-\, (\pi \cdot n) \pi_\mu \right] p_\nu}{S^2} \, , \nonumber\\
\left\{x_\mu , \pi_\nu \right\}_D \,&=&\,
\frac{\left[ \pi^2 n_\mu \,-\,( \pi
\cdot n )\pi_\mu \right] x_\nu }{S^2} \, , \nonumber \\
\left\{x_\mu , x_\nu \right\}_D \,&=&\,  \frac {x^2}{S^2}\,
\epsilon_{\mu \nu \gamma} S^\gamma \, , \\
\left\{p_\mu , p_\nu \right\}_D \,&=&\,
\frac {p^2}{S^2}  \,\epsilon_{\mu \nu \gamma} S^\gamma \, , \nonumber
\\
\left\{ x_\mu , p_\nu \right\}_D \,&=&\,
\frac {\left[ (x \cdot p) \epsilon_{\mu \nu \gamma}
S^\gamma  \,+\, S_\mu S_\nu \right]}{S^2}\nonumber.
\end{eqnarray}
Although rather complicated, the above brackets have some remarkable
features. Firstly, the Dirac brackets involving only the
extra variables $\left( n_\mu , \pi_\mu\right)$ are
exactly equal to the canonical Poisson brackets,
therefore we recover the canonical
spin algebra that we were searching, {\it i. e.}, one checks that

\begin{equation}
\left\{ S_\mu ,
S_\nu\right\}_D \,=\,\epsilon_{\mu \nu \gamma}S^\gamma.
\end{equation}
Furthermore, using (9) we deduce (for future comparison) :

\begin{equation}
\left\{ x_\mu , S_\nu \right\}_D \,=\,
\epsilon_{\mu \nu \gamma} x^\gamma \;\; ,\;\;
\left\{ p_\mu , S_\nu \right\}_D \,=\, \epsilon_{\mu \nu
\gamma}p^\gamma.
\end{equation}
By comparing (11) with the Poincar\`e algebra (1) we
notice that $S_\mu$ behave like the total angular
momentum  $J_\mu$ and (11) is nothing but the
Lorentz transformation law for the vectors
$x_\mu$ and $p_\mu$. Indeed, since the constraints (6) imply
$\epsilon_{\mu \nu \gamma}S^\nu p^\gamma\,=\,0$,
we analogously have from (8) that $\epsilon_{\mu \nu \gamma}S^\nu
x^\gamma\,=\,0$ and therefore the canonical
angular momentum $L_\mu \,=\,
\epsilon_{\mu \nu \gamma} x^\nu p^\gamma$ vanishes and we can identify
$S_\mu$ with $J_\mu$. Thus, the brackets
(10), (11) and the fact that $\left\{ p_\mu , p_\nu \right\}_D
\,\not = \, 0$ (see (9)) are simply telling us that the Poincar\`e
algebra (1) has been broken down to the Lorentz algebra
due to the lack of translational invariance of
the constraints (8). In the
next section we will show that the identification
between $S_\mu$ and $J_\mu$ and the absence of
translational invariance are not peculiar features of this
model but a consequence of the spin algebra and the
parallelness of $S_\mu$ and $p_\mu$.
Another important characteristic of the brackets (9)
to be noticed, is that they guarantee the
conservation of $p_\mu$ and $S_\mu$
$\left( {\dot  p}_\mu \,=\, \left\{p_\mu ,
H\right\}_D \,=\, 0 \,=\, {\dot S}_\mu\right)$ as
well as their gauge invariance. These conservation laws can
be used to show that ${\ddot x}_\mu \,=\, 0$
and therefore the model under study really corresponds
to a free particle, although the Dirac brackets
involving $x_\mu$ and $p_\mu$ are quite complicated.
So, we were able to reach the
goal of obtaining a free relativistic anyon
with canonical spin algebra.
\vskip .5cm
\section{A model independent analysis}

In this section, we will work from a more general point of view
without specifying how
the phase
space $(x_\mu , p_\nu )$ has to be extended to
include extra variables to describe the spin degree of
freedom, and that is precisely what we mean by a model
independent analysis.
Let us take our hamiltonian to be of the form
(5) but, since we are not going to specify the extended
phase space, we have, instead of four, N second
class constraints $\chi^A$ ($A \,=\,$ 1,...,N) where N will of course
depend on the number of variables of the extended phase
space. After the elimination of the
constraints $\chi^A$ by some Dirac bracket\footnote{ In this
section we relax the definition of the Dirac bracket in order
to include brackets used to eliminate second class constraints
which are not linearly independent.}
$\left\{ \,,\, \right\}^{\ast}$, the hamiltonian becomes simply

\begin{equation}
H \,=\, e\,\left( p^2 \,+\, m^2 \right) \,+\, \sigma \,
\left( S \cdot p \, +\,  \alpha \, m\right).
\end{equation}
Now we would like to examine the equations of
motion which come from $H$. For a generic phase space variable $q$ we
have,

\begin{equation}
{\dot q}\,=\,\left\{ q , H\right\}^{\ast} \,\equiv \,
\left\{ q , H \right\} \,-\, f_{A B} \left\{ q , \chi^A \right\}
\left\{\chi^B , H \right\}
\end{equation}
where $f_{AB}$ is some suitable function of the
phase space variables which certainly depends on specific features
of the extended phase space (see footnote).By using the fact that
in general we have $\,\{\chi^A,\phi_i\}\,=\,{c_i}^{Aj}\phi_j +
{d_{iB}}^A\chi^B\,$, where the functions $\,{c_i}^{Aj} , {d_{iB}}^A\,$
might depend on the details of the model studied, it is not
difficult to derive that, in a typical gauge where $e$ and $\sigma$
are set to a constant,
\begin{equation}
{\ddot x}_{\mu}\,=\,\sigma^2\epsilon_{\mu\nu\gamma}p^{\nu}S^{\gamma}+
h^i_{\mu}\phi_i
\end{equation}
where $h^i_{\mu}$ are functions of the phase space variables
which certainly depend on specific features of the extended phase
space.
Since $\varphi_{\mu}= \epsilon_{\mu\nu\gamma}p^{\nu}S^{\gamma}$ are
clearly independent of the first class constraints $\phi_i$ we
come to the important conclusion that, independent of the
specific details of the model, $p^{\mu}$ and $S^{\mu}$ must be
parallel,
i.e. $\varphi_{\mu}=0$, in order that the particle be free
(${\ddot x}_{\mu}=0$) and besides, we have to assume that the model
proposed
is such that the second term on the r.h.s. of (14) must independently
vanish.\footnote{ The reader can check that in order to
have $h^i_{\mu}=0$ it is sufficient, although not necessary, that
${c_i}^{Aj}=0$.}
Since $\phi_i \,(i=1,2)$ are the only first class constraints of
the theory the parallelness of
$p_\mu$ and $S_\mu$ must be a consequence of some of the second class
constraints $\chi^A$. Thus, we can use from now on the identification
(7) strongly. Furthermore,if we now impose that the particle
besides being free possess spin components $S_{\mu}$ which
are gauge invariant\footnote{ Note that this is a more severe
requirement than to simply impose that $S_{\mu}$ be observables.}
,i.e. $\,\{S_{\mu},\phi_i\}^* = 0\,$ and satisfy the canonical
spin algebra,

\begin{equation}
\left\{ S_\mu , S_\nu \right\}^{\ast} \,=\,
\epsilon_{\mu \nu \gamma} S^\gamma\quad ,
\end{equation}
we consequently have from (7) that
\begin{equation}
\left\{ S_\mu , p_\nu \right\}^{\ast} \,=\, \epsilon_{\mu \nu \gamma}
p^\gamma\quad.
\end{equation}
{}From (16) and (7) we obtain

\begin{equation}
\left\{p_{\mu},p_{\nu}\right\}^* = \frac{p^2}{S^2}
\epsilon_{\mu\nu\gamma}S^{\gamma}\quad ,
\end{equation}
now by using that $p_{\mu}$ is a vector, i.e.
$\,\{p_{\mu},J_{\nu}\}^*=\epsilon_{\mu\nu\gamma}p^{\gamma}\,$
where $\, J_{\mu}= \epsilon_{\mu\nu\gamma}x^{\nu}p^{\gamma} +
S_{\mu}\,$,
we can deduce the following expression

\begin{equation}
\epsilon_{\nu\gamma\beta}p^{\beta}\left\{p_{\mu},x^{\gamma}\right\}^*\
\frac{p^2}{S^2}\left( (S\cdot x) \eta_{\mu\nu} - x_{\mu}S_{\nu}\right)
\end{equation}
from which we have
$\, x_{\mu} = \left(S\cdot x /S\cdot p\right)p_{\mu}\,$
and thus $\,L_{\mu}=\epsilon_{\mu\nu\gamma}
x^{\nu}p^{\gamma}=0\,$ identically, consequently
$J_{\mu}$ and $S_{\mu}$ can be identified. Therefore it seems
that the only way out for the spin components of a free fractional
spinning particle to satisfy the usual angular momentum algebra is
to identify them with the total angular momentum $J_{\mu}$
which will certainly satisfy such algebra. Thus, we have seen that
the identification between
$S_{\mu}$ and $J_{\mu}$ as well as the lack of translational
invariance (see (17)) are not special features of the
last section model but a more general consequence of the
canonical spin algebra, Lorentz covariance and the requirement
that the particle be free. Besides the above general features,
we can also almost fully determine $\,\{x_{\mu},x_{\nu}\}^*\,$
and $\,\{x_{\mu},p_{\nu}\}^*\,$ from our basic assumptions as follows.
First notice that from the identification between $S_{\mu}$
and $J_{\mu}$ we have for the vector $x_{\mu}$:

\begin{equation}
\left\{ x_{\mu}, S_{\nu}\right\}^* \,=\, \epsilon_{\mu\nu\gamma}
S^{\gamma},
\end{equation}
since $x_{\mu}$ and $p_{\mu}$ are parallel like
$p_{\mu}$ and $S_{\mu}$ we can write $\, x_{\mu}=
\left(S\cdot x/S^2\right)S_{\mu}\,$, thus by using (19) and the
spin algebra (15) we obtain (compare with (9))

\begin{equation}
\left\{x_{\mu},x_{\nu}\right\}^* \, = \, \frac{x^2}{S^2}
\epsilon_{\mu\nu\gamma} S^{\gamma}
\quad .
\end{equation}
In order to fix $\,\{x_{\mu},p_{\nu}\}^*\,$ we have to work a little
more.
By use of the equality $\,\{x_{\gamma}, (S\cdot p)S_{\mu}\}^* =
\{x_{\gamma},S^2 p_{\mu}\}^*\,$ one can deduce

\begin{equation}
\left\{x_{\gamma},p_{\nu}\right\}^*{\cal P}^{\nu}_{\mu}\,
=\, \left(\frac{S\cdot p}{S^2}\right)\epsilon_{\gamma\mu\beta}
x^{\beta}\quad ,
\end{equation}
where $\,{\cal P}^{\nu}_{\mu} = \delta^{\nu}_{\mu} -
S^{\nu}S_{\mu}/S^2\,$
is a projection operator. Since $\{x_{\mu},p_{\nu}\}^*$
is a second rank tensor with 9 components, we can always decompose it
in terms of a 3 components vector $c_{\mu}$ and a 6 components
symmetric tensor $N_{\alpha\beta}$ without loss of
generality as follows

\begin{equation}
\left\{x_{\gamma},p_{\nu}\right\}^* = \epsilon_{\gamma\nu\mu}
c^{\mu} + N_{\gamma\nu}\quad .
\end{equation}
The quantities $c^{\mu}$ and $N_{\gamma\nu}$
can be partially determined from (21) and the
parallelness of $x^{\mu},p^{\mu}$ and $S^{\mu}$ such
that we end up, after some algebra, with

\begin{equation}
\left\{x_\mu , p_\nu \right\}^{\ast}
 \,=\, \frac{(p \cdot x)}{S^2}
\epsilon_{\mu\nu\gamma} S^\gamma \,+\, \frac{2\,b_\mu S_\nu }{S^2} \,-
\,
\frac{ (S \cdot b)S_\mu S_\nu }{(S^2)^2}
\end{equation}
where $b_\mu \,\equiv\, N_{\mu \nu} S^\nu $.

Since $S_\nu {\cal P}^\nu_\mu \,=\, 0$, we will have no further
informations about $b_\mu$ from
Eq. (21), but substituting (23) into
(18) we obtain
$\epsilon_{\mu\nu\gamma} p^\nu b^\gamma = 0$
and therefore $b_\gamma \,=\, f\,
S_\gamma$ where $f$ is an arbitrary
function of the extended phase space
variables. Back in (23) we get

\begin{equation}
\left\{ x_\mu , p_\nu \right\}^{\ast}
\,=\, \frac{(p \cdot x)}{S^2} \epsilon_{\mu\nu\gamma}
S^\gamma \,+\, f\, \frac{S_\mu S_\nu}{S^2}\quad.
\end{equation}
Although we cannot fully determine the function $f$ we can
obtain further constraints on it. Namely, all Jacobi identities
involving $x_\mu,\, p_\nu$ and $S_\gamma$ are satisfied if and only
if
$\,\left\{ f , S_\gamma \right\}^{\ast} \,=\, 0$, as can be
explicitly
checked. Moreover in order that the equations of motion for $x_\mu$
be
reproduced it is necessary that $f\,=\,1$ on shell. Since
there are many
candidates for $f$ which satisfy those constraints like, {\it e.g.},
$f\,=\,S\cdot p/\alpha m \, ; \, p^2/m^2$; etc.. . We think that the
specific form of $f$ depends on the specific model analysed, i.e.
depends
on how the phase space has been extended.

After having obtained (20),(24) and comparing with the results
of the model presented in section two, we see that all the brackets
involving $x_\mu , \, p_\mu$ and $S_\gamma$ completely agree if we
take
$f\,=\,1$ which is the simplest solution to our constraints.
Therefore we
are now able to recognize that the complicated brackets (9) are not
just an unpleasant feature of that specific model but a quite general
characteristic of the particles with fractional spin whose spin
components satisfy the canonical algebra. It is also clear that such
complications will go through the quantization of the model.

We conclude that for a free fractional spinning particle with the
canonical spin algebra, the constraints of the theory must
guarantee that $L_\mu\,=\,0$ and $S_\mu \propto p_\mu$, and in this
case we
willl be unavoidably faced with the rather complicated phase space
structure given in (17),(20), and (24).

In finishing this section we give a simpler way to derive such
results,
although less general, where we implement the above physical
conditions,
i.e. the parallelness of $S_{\mu},p_{\mu}$ and $x_{\mu}$
using  the constraints below

\begin{eqnarray}
\varphi_\mu &=& \epsilon_{\mu\nu\gamma}
p^\nu S^\gamma = \, 0, \nonumber \\
\tilde \varphi_\mu \,&=&\, \epsilon_{\mu\nu\gamma}
x^\nu S^\gamma =\, 0\quad .
\end{eqnarray}
It is easy to show that those constraints are all second class, thus
they
could in principle be eliminated by a Dirac bracket, however this
is not possible \cite{Henneaux} since the matrix of the
Poisson brackets between the constraints has no
inverse due to the fact that not all constraints $\varphi_\mu$ and
$\tilde
\varphi_\mu$ are linearly independent. Nevertheless we can eliminate
them
iteractively by defining a kind of Dirac bracket as follows:
\cite{Henneaux}
first we eliminate, {\it e.g.}, $\varphi_\mu$ through the bracket
$\left\{
\,,\, \right\}^{\dagger}$,

\begin{equation}
\left\{ A , B \right\}^{\dagger} \,=\, \left\{ A , B \right\} \,+\,
\frac{p_\gamma}{p^2(S \cdot p)} \epsilon^{\gamma \delta \beta}
\left\{ A ,
\varphi_\delta , \right\} \left\{ \varphi_\beta , B \right\},
\end{equation}
and finally we eliminate $\tilde \varphi_\mu$ through $\left\{ \,,\,
\right\}^{\ast}$:

\begin{equation}
\left\{ A , B \right\}^{\ast} \,=\, \left\{ A , B \right\}^{\dagger}
\,+\, \frac{p^2}{S^2(S \cdot p)} p_\gamma
\epsilon_{\mu\nu\gamma} \left\{ A , {\tilde
\varphi}_\mu \right\}^{\dagger} \left\{ {\tilde \varphi}_\nu , B
\right\}^{\dagger}.
\end{equation}
The reader can check that after the elimination of $\varphi_\mu$ and
${\tilde \varphi}_\mu$, the brackets involving $S_\mu, \, p_\mu$ and
$x_\mu$ will match with the general results obtained in this section
for
the case $f\,=\,1$.

\section {Conclusion and Summary}

Based on some mild assumptions like Lorentz covariance and the fact
that a fractional spinning particle in $2+1$ dimensions obeys,
analogous to the spin 1/2 and spin 1 particles \cite{Jackiw}, an
equation
of the form $S\cdot P + \alpha m = 0$, we have shown in this work
that some peculiar features of some  point-particle models recently
proposed
in the literature  can be understood from a rather general point of
view. The
first of such features is that, irrespective of the spin algebra, we
can
only have a free (${\ddot x}_{\mu}=0$) fractional spinning particle
whenever spin and momentum are parallel, as an example of such
rule we can take two models defined in \cite{Plyuschay}; for the
first model, defined in eqs. (2.12)-(2.14) of \cite{Plyuschay},
$S_{\mu}$
and $p_{\mu}$ are parallel whereas this is not the case for
the second one, worked out in the fourth section of the
same reference, and indeed we have  respectively ${\ddot x}_{\mu}=0$
and ${\ddot x}_{\mu}\ne 0$.
Furthermore, if the particle besides of being free is required
to have spin components which are gauge invariant and satisfy
the canonical angular momentum algebra, both in the strong sense,
then the algebra of the quantities $S_{\mu},p_{\nu},$ and the
coordinates $x_{\mu}$ can be, as we have shown, almost completely
fixed independently of the details of the
extended phase space introduced to describe the spin degree
of freedom. It has been shown in particular that our basic
assumptions lead to theories where the Poincar\`e symmetry
is broken
down to the Lorentz one and the spin components will
necessarily play the role of the Lorentz generators. These general
predictions have been confirmed in an example given in the
first section of this article. Moreover we have also seen that
the canonical structure of the variables $x_{\mu}$
and $p_{\nu}$ will be quite complicated which will clearly
pose some difficulties in the quantization program.

Finally, it should be noticed that it is no yet clear
in the literature (see \cite{Jackiw} and comments in \cite{Forte} and
references
therein) whether fractional spinning particles in $2+1$ dimensions
can be really formulated as free relativistic field theories and,
although
our assumptions can be softened, we think that the rather complicated
canonical structure and the lack of translational invariance found
here can be seen as indications of this fact.
\vskip .8cm

\centerline {\bf{ Acknowledgements}}
Both authors are partially supported by CNPq. Financial support
of FAPESP (contract No. 93/1476-3 and No. 93/0595-9) is also
acknowledged.

\end{document}